\documentclass[preprint,showpacs,aps]{revtex4}
 \draft
\begin{document}

\title{Atomistic simulations of the incipient ferroelectric KTaO$_3$}
\author{A.R. Akbarzadeh$^{1}$, L. Bellaiche$^{1}$, Kevin Leung$^{2}$, Jorge \'I\~niguez $^{3,4,5}$ and David Vanderbilt$^{3}$}
\address{$^{1}$Physics Department, University of Arkansas, Fayetteville, AR 72701, Arkansas, USA}
\address{$^{2}$Sandia National Laboratories, Mail Stop 1421,Albuquerque, New Mexico 87185, USA}
\address{$^{3}$Department of Physics and Astronomy, Rutgers University, Piscataway, New Jersey 08854-8019, USA}
\address{$^{4}$NIST Center for Neutron Research, National Institute of Standards and Technology, Gaithersburg, Maryland 20899-8562}
\address{$^{5}$Department of Materials Science and Engineering,
University of Maryland, College Park, Maryland 20742-2115}

\date{\today}


\begin{abstract}
A parameterized effective Hamiltonian approach is used to
investigate KTaO$_3$. We find that the experimentally observed anomalous
dielectric response of this incipient ferroelectric is well reproduced by this
approach, once
quantum effects are accounted for.  Quantum fluctuations
suppress the paraelectric--to--ferroelectric phase transition; it
is unnecessary to introduce defects to explain the dielectric
behavior.  The
resulting quantum-induced local structure exhibits off-center
atomic displacements that display {\sl longitudinal, needle-like correlations}
extending a few lattice constants
\end{abstract}

\narrowtext \maketitle
\marginparwidth 2.7in
\marginparsep 0.5in

Numerous experimental and theoretical studies have been carried
out on the perovskite KTaO$_3$ over the last forty years (see,
e.g,
Refs~\cite{Kvyatkovskii,salce,Wemple,singhKTaO3,Kevin,Tinte,ref7,ref8,raman,ref6,Ang,comes}
and references therein), making this material one of the
most-studied ``incipient ferrolectrics.'' The main reason for this
interest is that the dielectric constant of KTaO$_3$ increases
continuously with decreasing temperature down to $\sim$10K, but
then saturates to a plateau at a large value ($\simeq$ 4000) at
lower temperatures while remaining paraelectric and cubic all the
way down to zero Kelvin \cite{salce,Wemple}. These anomalous
low-temperature features are usually thought to be caused by the
suppression of a paraelectric--to--ferroelectric phase transition
by zero-point quantum fluctuations \cite{salce,Wemple} (hence the
name ``incipient ferroelectric'' or ``quantum paraelectric'' used
to describe KTaO$_3$ and other materials, such as SrTiO$_3$,
exhibiting similar unusual dielectric and structural properties).
Surprisingly, this generally-accepted picture is apparently {\it
not} supported by various first-principles calculations, using
density-functional theory (DFT) either in its local-density
approximation (LDA)\cite{Kohn} or generalized-gradient
approximation (GGA)\cite{pw91,pbe} form, since these simulations
all predict that KTaO$_3$ should be paraelectric at $T$=0 even
when neglecting zero-point motion \cite{singhKTaO3,Kevin,Tinte}.
This raises the possibility that LDA and GGA are not accurate
enough to adequately reproduce the qualitative properties of
incipient ferroelectrics. An alternate explanation for this
discrepancy between first-principles calculations and experiments
is that the simulations assume a perfect material while real
samples may contain defects such as oxygen vacancies and Fe$^{+3}$
ions \cite{salce,Wemple,ref7,ref8} that might lead to the observed
anomalous properties of KTaO$_3$. In fact, the interpretations of
various experiments \cite{raman,ref6} still remain controversial
as to whether they are attributable to extrinsic effects (i.e.,
defects-induced) or intrinsic off-center atomic displacements.
Furthermore, while previous studies invoke the existence of {\it
ferroelectric microregions} inside the {\it
macroscopically-paraelectric} KTaO$_3$ system to explain some of
its properties \cite{raman,Ang}, there has never been any {\it
direct} determination of the size and shape of these proposed
polar regions, to the best of our knowledge. For instance, the
pioneering work of Ref.~\cite{raman} made several assumptions in
their analysis of low-temperature Raman spectra -- such as
isotropy of these microregions -- to extract a characteristic size
$\simeq$ 16 \AA\ for these polar regions.

In this Letter, we use large-scale atomistic simulations to shed
light on the aforementioned long-standing problems. We report
calculations on KTaO$_3$ using a parameterized effective
Hamiltonian approach. Our main findings are that (i) LDA and GGA
are indeed not accurate enough to reproduce the observed anomalous
properties of KTaO$_3$, even qualitatively; (ii) these properties
{\it can} be understood {\it without the need of introducing
defects}, if quantum fluctuations are present to suppress the
paraelectric--to--ferroelectric transition; (iii) the
low-temperature local structure of KTaO$_3$ is characterized by
off-center atomic displacements that are {\it longitudinally
correlated, in a needle-like (and thus anisotropic) way, with a
correlation length spanning a few 5-atom unit cells}.

We use the effective Hamiltonian ($H_{\rm eff}$) approach developed in
Ref.~\cite{ZhongDavid} to investigate finite-temperature properties
of KTaO$_3$.  Within this approach, the total energy $E_{\rm tot}$ is a function of
three types of local degrees of freedom:
(1) the {\bf u$_i$} (B-site centered) local soft-mode amplitude
in each $i$ 5-atom cell, describing the local polarization in each cell;
(2) the {\bf ${v_i}$} (A-site centered) inhomogeneous strain variables;
and (3) the homogeneous strain tensor. $E_{\rm tot}$ contains
 18 parameters and 5 different contributions: a
 local-mode self energy, a long-range dipole-dipole interaction, a short-range interaction between local modes,
 an elastic energy, and an interaction between the local modes and strains \cite{ZhongDavid}.
 This effective  Hamiltonian approach has been successfully used to model, understand, and design
 ferroelectric perovskites (see Refs.~\cite{ZhongDavid,DavidReview,LaurentReview,jorge,kevin2} and references therein).
$E_{\rm tot}$ is used in two different kinds of
Monte-Carlo (MC)
 simulations: classical Monte Carlo (CMC) \cite{Metropolis},
which does not take into
 account zero-point phonon vibrations,
 and path-integral quantum Monte Carlo (PI-QMC) \cite{jorge,zhong,ceperley}
 which includes purely quantum-mechanical zero-point motion.
Consequently, comparing the results of these two
 different Monte-Carlo techniques allows a precise determination of
 quantum effects on macroscopic and microscopic properties of perovskites.
 12$\times$12$\times$12 KTaO$_3$ supercells (corresponding to 8,640 atoms)
 are used in all Monte-Carlo simulations. We typically perform
 30,000 MC sweeps to thermalize the system and
 70,000 more to compute averages,
 except at low temperatures in PI-QMC where more statistics is needed.
 For example, we use 180,000 MC sweeps for
 thermalization and 240,000 sweeps at 3K to accurately predict the dielectric response.
 (Note that we are not aware of any previous work reporting the
 dielectric response computed using PI-QMC)

In PI-QMC, each 5-atom cell interacts with its images at
neighboring imaginary times through a spring-like potential
(mimicking the zero-point phonon vibrations), while all the 5-atom
cells interact with each other at
 the same imaginary time through the internal potential associated with $E_{\rm tot}$.
The product $TP$, where $T$ is the simulated temperature and $P$
is the number of imaginary time slices (Trotter number), controls
the accuracy of the PI-QMC calculation.  In all our simulations we
use $TP$=600, which we find leads to sufficiently converged
results. Outputs of the PI-QMC simulations thus contain
local-modes {\bf u$_i$}(t), where $i$ indexes the 5-atom unit
cells of the studied supercell while the imaginary time $t$ ranges
between 1 and $P$. Note that CMC simulations
 can be thought of as corresponding to $P=1$, so that they do not yield
 imaginary-time-dependent outputs.

 Figure 1(a) shows the $\chi_{33}$ dielectric susceptibility -- where
 the index $3$ refers to the [001] pseudo-cubic direction --
as predicted by the $H_{\rm eff}$ approach, {\it with all its parameters being
 derived from LDA calculations on small supercells of  KTaO$_{3}$ at
 its experimental lattice constant}. (Technical details
 of these LDA calculations are similar to those of Ref.~\cite{Kevin}).
It can be clearly seen that CMC calculations yield
a $\chi_{33}$ that is continuously increasing as the temperature is
decreasing down to nearly zero Kelvin.  Turning on quantum
effects leads to the appearance of a plateau below $\sim$100K
with a value of $\sim$100 for the dielectric constant. These CMC and
PI-QMC simulations {\it both} predict a cubic paraelectric ground state.
A plateau for the dielectric response has indeed been
experimentally observed in KTaO$_3$ \cite{salce,Wemple}, but reaching
a much higher dielectric constant ($\simeq$ 4,000) and over a much narrower
temperature range (i.e., below 10K) than in Fig.~1(a).

In view of these two discrepancies, we have experimented with making
minor adjustments in the LDA-fitted parameters in the hope of obtaining
better agreement with experimental data.  We have found that this can
be done by adjusting just one of the 18 parameters, namely,
the parameter denoted $\kappa_2$ in Ref.\cite{ZhongDavid},
which describes the harmonic part of the local-mode self-energy.
(In our model, reducing $\kappa_2$ favors ferroelectricity
with respect to paraelectricity since it leads to a decrease of the
zone-center transverse optical frequency by weakening short-range repulsions).
 Figure 1b shows that decreasing this single $\kappa_2$ parameter by $\sim$18 \% from its LDA value of
 0.0866 a.u. (atomic units) leads to reasonable agreement between our PI-QMC simulations and measurements,
not only for the value of the dielectric constant plateau, but
also at temperatures above 10K.

Furthermore, this modified $\kappa_2$ also
results in a dramatic difference between the two kinds
of Monte-Carlo calculations. CMC simulations yield a {\it
ferroelectric} rhombohedral ground-state.  The corresponding
Curie temperature is around 30K, as evidenced by the peak in
dielectric response displayed in Fig.~1(b). On the other hand, PI-QMC predicts a
{\it paraelectric} ground state. In other words, {\it quantum
effects suppress the paraelectric--to--ferroelectric phase
transition}, which is consistent with the accepted picture \cite{salce,Wemple}.
Figures 1(a-b) thus (i) reveal that extrinsic
defects (such as impurities or vacancies), which have been proposed
to be responsible for the anomalous properties of KTaO$_3$
\cite{salce,Wemple,ref7,ref8}, are {\it not} needed to reproduce
the experimental behavior of this material; and (ii) strongly suggest
that, unlike in {\sl strongly ferroelectric}
perovskites\cite{DavidReview,LaurentReview}, the LDA is not accurate enough for
simulating KTaO$_3$.

As for GGA, Tinte {\sl et al.}~\cite{Tinte} report zone-center
optical frequencies in cubic KTaO$_3$ that are all positive and
very close to the LDA values. Consequently, according to Fig
1(a-b), we can conclude that a GGA effective Hamiltonian would not
provide a significant improvement over our LDA one, and will also
fail in reproducing experimental results. This may make KTaO$_3$ a
useful test-case for the development of new functionals within DFT
or other {\it ab-initio} methods.

We now analyze the {\it microscopic} local structure of KTaO$_3$
at low temperature.  Figure~2 depicts the
magnitude of the local modes {\bf u$_i$} inside each $i$ 5-atom
 cell {\it versus} the angle that these local modes make with the pseudo-cubic [100]
direction, as obtained from a $T$=3K snapshot among the thermally
equilibrated Monte-Carlo configurations using $E_{\rm tot}$ with
the modified $\kappa_2$. (The magnitude of the local mode is
directly proportional to the magnitude of the local polarization,
e.g., ${\left|{\bf u}\right|}=$ 0.006 and 0.026 a.u. correspond to
a local polarization $\simeq$ 0.0583 and 0.253 ${C}/{m^2}$,
respectively). Figure~2(a) displays the CMC results, while
Fig.~2(b) corresponds to PI-QMC \cite{footnoteJorge}. Comparing
Figs.~2(a) and ~2(b) reveals how quantum effects affect the
microscopic structure of KTaO$_3$: the local polarizations go from
all lying close to the [111] direction (corresponding to an angle
$\simeq$ $54^\circ$) and having a relatively large magnitude, to
being heavily-scattered in direction and having a much smaller but
non-zero magnitude. The fact that KTaO$_3$ is predicted to exhibit
non-zero local dipoles, even when quantum fluctuations are
accounted for,  is consistent with the first-order lines observed
to appear in Raman spectra which are forbidden in the ideal cubic
perovskite structure \cite{raman,ref6}. Furthermore, an inspection
of Fig.~2(b) does {\it not} reveal any obvious polar microregions.
For instance, our quantum-statistical results do not show the
local-mode distributions breaking up into clusters centered along
$\langle 111 \rangle$ directions (i.e., angles of $\simeq$ $54$
and/or $125^\circ$) as would be expected for such polar
microregions.

To gain further insight into the local structure of KTaO$_3$, we
decided to compute an additional set of coefficients defined as
\begin{equation}
\theta_{\mu}({\bf r}) =\frac{3}{N} \sum_{i=1}^N
\frac{u_{i,\mu}~u_{i+{\bf r},\mu}}{\left|{\bf
u_i}\right|\left|{\bf u}_{i+{\bf r}}\right|} ~~.
\label{eq:thetamu}
\end{equation}
Here $\mu$ denotes the $x$, $y$, or $z$ Cartesian axis chosen along
the [100], [010] or [001] cubic directions, respectively.
The index $i$ runs over all the $N$ B-sites; $u_{i,\mu}$ and
$u_{i+{\bf r},\mu}$ are the $\mu$ components of the local
modes in cell $i$, and in the cell centered at a distance
${\bf r}$ from cell $i$, respectively. The case in
which the local dipoles all have the same (non-zero) magnitude and
are all aligned along a given $\langle 111 \rangle$ direction yields a value
of 1 for $\theta_\mu({\bf r})$, for any ${\bf r}$ and for any
$\mu$. This case corresponds to a ferroelectric rhombohedral state
having identical local and average structures. On the other hand,
the other limiting case
--- for which neighbors at a distance ${\bf r}$ do not exhibit any correlation
between the $\mu$-components of their local modes
--- is associated with a zero value for $\theta_{\mu}({\bf r})$.

Figure 3 depicts $\theta_x({\bf r})$ (i.e., $\mu=x$) for $\bf r$
lying in the $x$-$y$ plane. The results correspond to one snapshot
of a thermally equilibrated Monte Carlo configuration at $T$=3K,
using the $H_{\rm eff}$ with the modified $\kappa_2$. Panels (a)
and (b) correspond to CMC and PI-QMC simulations respectively. One
can see that CMC technique leads to a $\theta_x({\bf r})$ close to
unity for any ${\bf r}$, and thus generates a macroscopically- and
microscopically-ferroelectric rhombohedral structure, as
consistent with Fig.~2(a).

On the other hand, PI-QMC simulations give a more complex
behavior for $\theta_x({\bf r})$ at low temperature. One can see that
the $x$ components of the local modes are {\it longitudinally}
correlated in a {\it needle-like} fashion: $\theta_x({\bf r})$
adopts large values only when ${\bf r}$ is along the [100]
direction.  These values decreasing as the magnitude of ${\bf
r}$ increases.
(The same result is obtained for all symmetry related cases, e.g.,
for $\theta_x({\bf r})$ in the $x$-$z$ plane, etc.) Figure~2(b)
further reveals that $\theta_x({\bf r}) \simeq 0.5$ for neighbors
at a distance of $\pm 2a$ (where $a \simeq 4\,$\AA\ is the cubic
lattice constant) along the $x$ axis.  This is in good agreement
with the characteristic size of $16\,$\AA\ extracted from
low-temperature Raman spectra of KTaO$_3$ \cite{raman}. On the
other hand, our simulations go against the hypothesis of isotropic
correlation made in ref.~\cite{raman}. The longitudinal
needle-like correlations depicted in Fig.~3(b)  have also been
predicted to occur in classical ferroelectrics just above the
paraelectric--to--ferroelectric transition
temperature~\cite{DavidFE1998}.
In fact, they are pretransitional effects that are probably common to
most ferroelectric perovskites, the peculiarity of
quantum paraelectric KTaO$_3$ being that the phase transition does not
actually occur.
Finally, note that these needle-like correlations are
consistent with the peculiar diffuse X-ray
scattering observed in Ref.~\cite{comes}.

In summary, we have performed large-scale atomistic simulations to
investigate the (defect-free) incipient
ferroelectric KTaO$_3$ system using a parameterized effective-Hamiltonian
approach.  The effect of quantum-mechanical zero-point
motion is investigated by comparing the results of classical
and path-integral Monte Carlo simulations.
We find that the fitting of all the
$H_{\rm eff}$ parameters within LDA yields a theoretical
dielectric constant that is in poor quantitative agreement with
experiment, strongly suggesting that LDA is inadequate for
this material.
Results in the literature also indicate that GGA will not
improve the LDA result.
On the other hand, a small modification of a single parameter in
$H_{\rm eff}$ from its LDA value is enough to obtain
reasonable agreement between theory and experiment for the
dielectric constant over a wide temperature range.
This modified $H_{\rm eff}$ leads to the
predictions that (i) KTaO$_3$ is ferroelectric classically, but
becomes paraelectric once zero-point phonon vibrations are included,
and (2) the quantum-induced
local structure of KTaO$_3$ is characterized by non-zero local dipoles
that have longitudinal, needle-like correlations with a correlation
length spanning a few unit cells.

We thank D. Ceperley, B. Dkhil, M. Itoh, J.M. Kiat, W. Kleemann,
J. Kohanoff, I. Kornev, S.A. Prosandeev, G. Samara, J. Shumway,
and E. Venturini for useful discussions. This work is supported by
Office of Naval Research Grants N00014-01-1-0365 (CPD),
N00014-01-1-0600 and N00014-97-1-0048, National Science Foundation
Grant DMR-9983678 and Department of Energy Contract
DE-AC04-94AL85000 at Sandia National Laboratories.

\newpage

\begin{figure}
\caption{$\chi_{33}$ dielectric susceptibility of KTaO$_3$ as a
function of temperature $T$.  (a) Results for LDA-fitted
$H_{\rm eff}$ parameters.  (b) Results for modified set of parameters.
Solid circles and stars correspond to PI-QMC and  CMC results respectively.
Dashed and dotted lines
represent experimental data from Refs.~\protect\cite{salce} and \protect\cite{Wemple},
respectively.
Solid line shows the fit of the PI-QMC results by a Barrett
relation $A/[(T_1/2)\,\coth(T_1/2T)-T_0]$ \protect\cite{Barrett}, with $A=27000$,
$T_1=72$K and $T_0=29$K. Note that our CMC simulations
yield a paraelectric--to--ferroelectric transition around 30K, which
provides a numerical proof for the concept of classical Curie temperature
given to $T_0$ in the Barrett relation.}
\end{figure}

\begin{figure}
\caption{Magnitude of local modes of KTaO$_3$ at T=3K  {\it
versus} the angle that these modes make with respect to the [100]
pseudo-cubic direction. The modified set of $H_{\rm eff}$
parameters is used.}
 \end{figure}

\begin{figure}
\caption{Correlation function $\theta_x({\bf r})$ of
Eq.~(\protect\ref{eq:thetamu}) for KTaO$_3$ plotted in the $x$-$y$
plane for a 12$\times$12$\times$12 simulation at $T$=3K. (a) CMC
results; (b) PI-QMC results.  Each small square represents one
lattice B site; the origin lies at the center.  The modified set
of $H_{\rm eff}$ parameters is used. Note that the color scales
are different in the two panels.}
\end{figure}

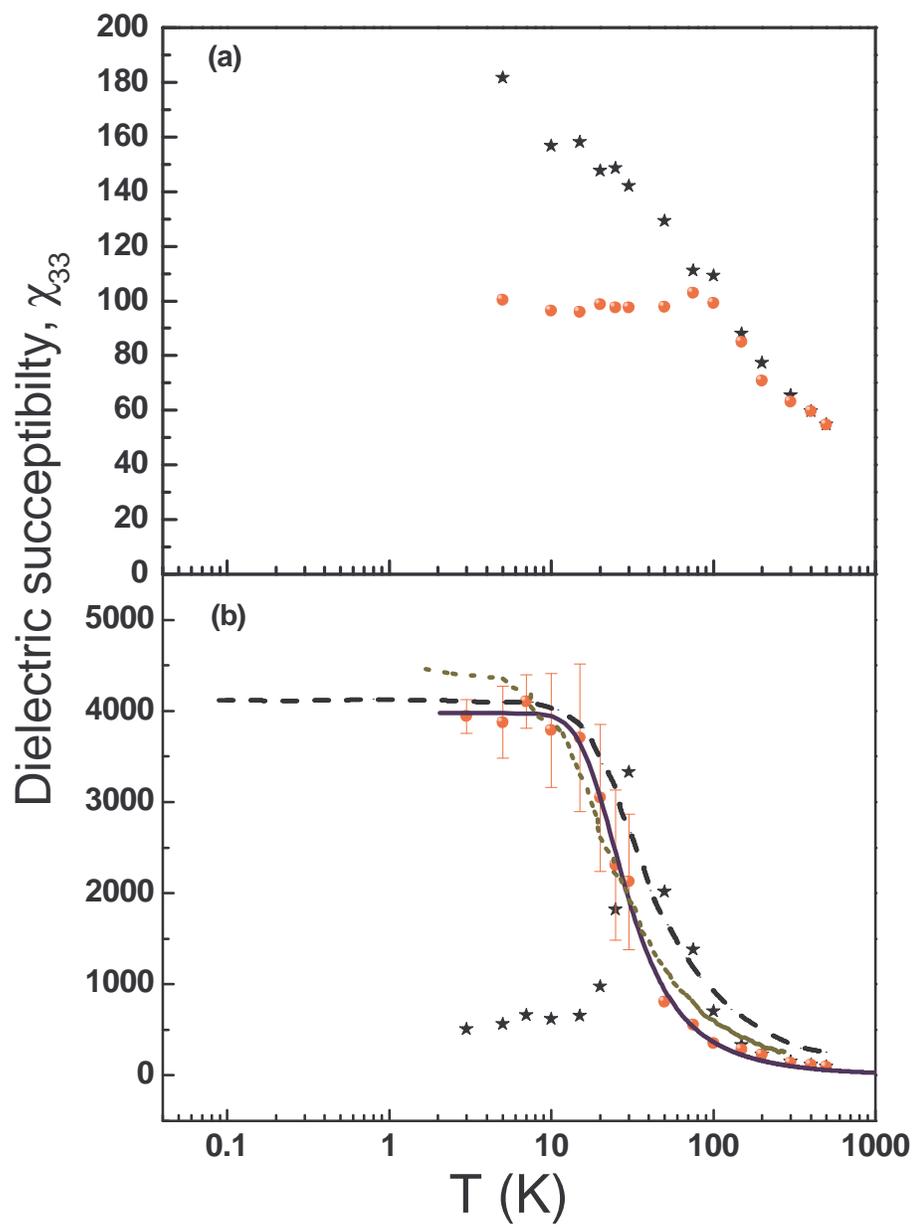
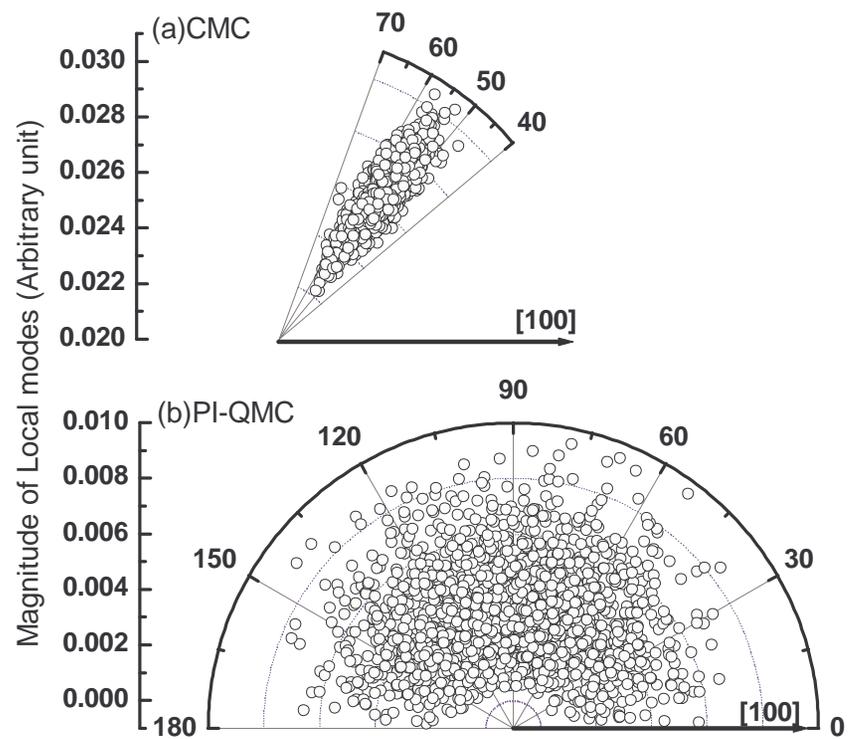
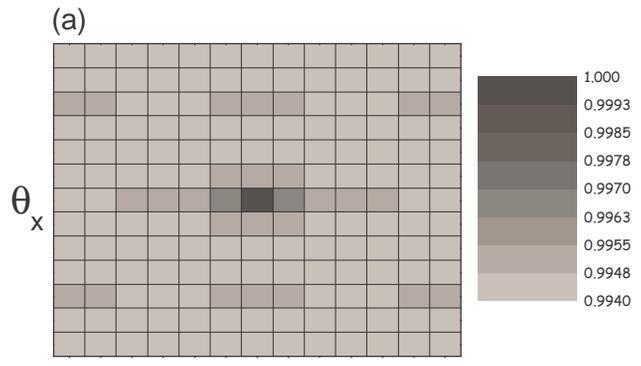
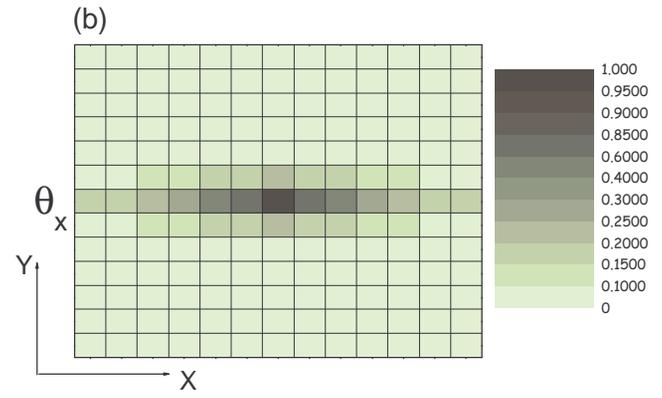
\end{document}